\documentclass[fleqn,twoside]{article}
\usepackage{espcrc2}
\usepackage{graphicx}
\usepackage{epsfig}
\usepackage[figuresright]{rotating}

\newcommand{\AmS}{{\protect\the\textfont2
  A\kern-.1667em\lower.5ex\hbox{M}\kern-.125emS}}
\title{The measurement of $\theta_{13}$ and $\delta$: 
the role of the uncertainties on the solar and atmospheric parameters}

\author{D. Meloni\address[Roma]{INFN, Sezione di Roma e Dip. di Fisica, 
Univ. di Roma ``La Sapienza'', P.le A. Moro 2, I-00185 Roma, Italy}}
       
\begin{document}

\begin{abstract}
In this talk we show how the errors on solar and atmospheric parameters
  affect the measurement of the unknown PMNS parameters 
  $\theta_{13}$ and $\delta$ at future LBL facilities. Performing 
  three parameters fits in $\theta_{13}$, $\delta$ and, in turn, one of the 
  atmospheric or solar parameters, 
  we show that 
  present uncertainties on $\theta_{23}$ and $\Delta m^2_{23}$  
  worsen significantly the precision on ($\theta_{13}$,$\delta$) whereas the
  solar sector does not introduce further uncertainties.
  A precision on the atmospheric parameters similar to what expected at T2K-I 
  is necessary to improve the sensitivities to $\theta_{13}$ and $\delta$.

\vspace{1pc}

\end{abstract}

\maketitle

\section{Motivations}

The atmospheric and solar sector of the PMNS leptonic mixing matrix have been measured 
with quite good resolution by SK, SNO and KamLand. These experiments measure two angles, 
$\theta_{12}$ and $\theta_{23}$, and two mass differences, $\Delta m^2_{12}$ and 
$\Delta m^2_{23}$. On the other hand, only an upper bound exists on 
the other angle, $\theta_{13}$, and the CP-violating phase $\delta$ is completely unknown \cite{lisi}. 
Two additional discrete unknowns are the sign of the atmospheric mass difference and 
the $\theta_{23}$-octant (if $\theta_{23} \neq 45^\circ$).

The strong correlations between $\theta_{13}$ and 
$\delta$ and the presence of parametric degeneracies in the 
($\theta_{13},\delta$) parameter space \cite{rigolin} (the so-called {\it clones}), make 
the simultaneous measurement of the two variables extremely difficult. 
In the literature, this has been 
normally studied considering the solar and atmospheric mixing parameters as external quantities 
fixed to their best fit values 
(see for example Ref.~\cite{Apollonio:2002en} and refs. therein) . However, the experimental uncertainties on 
these parameters can in principle affect the measurement of the unknowns, and it seems important 
to perform an analysis that goes beyond the two-parameters fits presented in the literature.

In this talk we therefore present the impact that ``solar'' (i.e. $\theta_{12}$ 
and $\Delta m^2_{12}$) and ``atmospheric'' (i.e. $\theta_{23}$ and $\Delta m^2_{23}$) parameters 
uncertainties have on the measurement of $\theta_{13}$ and $\delta$, studying
their effects at  
three of the many proposed setups: 
the 4 MWatt SPL Super-Beam~\cite{Gomez-Cadenas:2001eu}, a $\gamma \sim 100$ 
$\beta$-Beam~\cite{Bouchez:2003fy} 
and the CERN-based 50 GeV Neutrino Factory 
(considering both the ``golden'' \cite{Cervera:2000kp} 
and ``silver'' \cite{Donini:2002rm} channels). 
\section{The strategy}

With the aim of understanding how any single parameter affects the measurements of $\theta_{13}$ and $\delta$, we
performe a series of three-parameters fits (taking $x$=$\theta_{12},\Delta m^2_{12},\theta_{23}$ 
and $\Delta m^2_{23}$ in turn as the third fitting variable) to be compared with standard 
two-parameters fits in $\theta_{13}$ and $\delta$. For each transition channel 
$\nu_\alpha \to \nu_\beta$, sign of $\Delta m^2_{23}$ ($s_{atm}=\pm 1$ for positive and negative values
respectively) and
octant of $\theta_{23}$ ($s_{oct}=\pm 1$ for $\theta_{23} > ~$or$~ <45^\circ$ respectively), 
we build a $\chi^2$ function :
\begin{eqnarray}
\left [ \chi^2(\theta_{13}, \delta, x) \right ]_{\alpha\beta} = 
\sum_\pm \left[\frac{
N^\pm_{\alpha\beta} (\vec{g})-
N^\pm_{\alpha\beta} (\vec{t} )
}{\delta N^\pm_{\alpha\beta}}\right]^2 \, ,
\label{chi2}
\end{eqnarray}
where $\vec{g}=(\theta_{13}, \delta, x; s_{atm}, s_{oct})$ and 
$\vec{t}=(\bar \theta_{13}, \bar \delta, \bar x;  \bar s_{atm}, \bar s_{oct})$ are vectors of {\it guessed} and
{\it true} (that is, chosen by Nature) parameters.
$\pm$ refers to neutrinos 
or antineutrinos and $N^\pm_{\alpha\beta}$ is the number of charged leptons $l^\pm_\beta$ observed in the detector 
for a $\nu_\alpha (\bar \nu_\alpha)$ beam. 
The error $\delta N^\pm_{\alpha\beta}$ on the sample takes into account 
the statistical error on $N^\pm_{\alpha\beta}$ as well as 
the sum of beam and detector backgrounds and the total systematic 
error (see \cite{Meloni:2004ee} for details).
The three-parameters $\chi^2$ function defines a three-dimensional 90\% CL contour that is 
eventually projected onto the ($\theta_{13},\delta$) plane to perform a direct comparison with 
the standard two-parameters 90\% CL contours for the considered
setups\cite{altridon}. 

\section{The results}

By comparing the results at the three, very different, facilities, we deduce 
that the impact of the current atmospheric parameters uncertainties is a common problem
that future experiments looking for $\theta_{13}$ and $\delta$ will have 
to face.  

\begin{figure}[htb]
\vspace{-0.25cm}
\epsfxsize6.5cm\epsffile{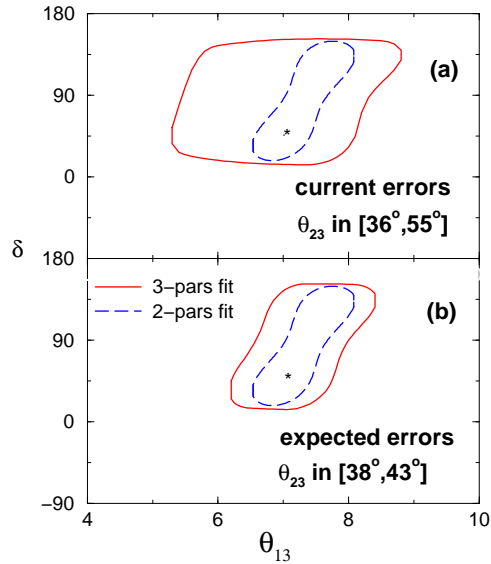}
\vspace{-0.85cm}
\caption{\it Comparison of the projection of three-parameters 90\% CL contours onto the ($\theta_{13}$, $\delta$) plane 
(solid lines) with the corresponding two-parameters 90 \% CL contours (dashed lines) 
after a 10 years run at the $\beta$-Beam. Star represent the projection of the
input point ($\bar \theta_{13}$, $\bar \delta$, $\bar \theta_{23}$ )=($7^\circ$,
$45^\circ$, $40^\circ$). See \cite{Meloni:2004ee} for further details.}
\label{fig:bb}
\end{figure}

Especially for larger values of $\bar \theta_{13}$ (and almost every
value of $\bar \delta$), the present uncertainties on the atmospheric parameters
are large enough to modify in a significant way the results of 
two-parameters fits, resulting in a lost of precision on $\theta_{13}$ and 
$\delta$. We have noticed that this is mainly due to 
the wide displacements of the clones, which are free to move in the 
multi-dimensional manifold to arrange for a lower $\chi^2$.
An example at the $\beta$-Beam facility is shown in plot (a) 
of Fig.\ref{fig:bb} in which the output of
our statistical approach (for $x=\theta_{23}$ and the {\it true} solution only, 
$s_{atm}=\bar s_{atm}$ and $s_{oct}=\bar s_{oct}$) is compared with a usual
two-parameters fit.

The situation is quite different if the precision on the atmospheric 
parameters is improved. This could be achieved by T2K-I experiment 
\cite{Itow:2001ee} and by the SPL Super-Beam itself. Using their expected errors
on $\theta_{23}$ and $\Delta m^2_{23}$, we have observed that in much 
of the ($\theta_{13}$, $\delta$) parameter space there is a general improvement on
the precision of these parameters (see plot (b) in Fig.\ref{fig:bb})
but some extra clones
are still present in three-parameters fits that were absent in the 
two-parameters analysis. 
This is a clear indication of the 
fact that the problem we are addressing 
must be seriously taken into account when envisaging future facilities to look for
$\theta_{13}$ and $\delta$, even in the case of reduced uncertainties on 
$\theta_{23}$ and $\Delta m^2_{23}$.

As a final remark, we stress that the impact of solar parameters 
uncertainties on the measurement of ($\theta_{13},\delta$)
is negligible, at least above the verge of the $\theta_{13}$-sensitivity for the
considered facilities.

\end{document}